\documentclass[twocolumn,prl,superscriptaddress]{revtex4-1}
\usepackage{amsmath,amssymb,mathrsfs}
\usepackage{natbib}
\usepackage{subfigure}
\usepackage{tabularx}
\usepackage{epsfig}
\usepackage{longtable}
\usepackage{amsfonts}
\usepackage{rotating}
\usepackage{bbold}
\usepackage{hhline}
\usepackage{braket}
\usepackage{txfonts, comment}
\usepackage{multirow}
\usepackage{appendix}
\usepackage[unicode=true,bookmarks=true,bookmarksnumbered=false,bookmarksopen=false,breaklinks=false,pdfborder={0 0 1},backref=false,colorlinks=true]{hyperref}

\hypersetup{linkcolor=magenta,urlcolor=blue,citecolor=blue,pdfstartview={FitH},hyperfootnotes=false,unicode=true}

\def\be{\begin{equation}}
\def\ee{\end{equation}}
\def\bea{\begin{eqnarray}}
\def\eea{\end{eqnarray}}

\begin{document}

\title{Schr\"{o}dinger-picture formulation of a scalar quantum field driven by white noise}

\author{Pei Wang}
\affiliation{Department of Physics, Zhejiang Normal University, Jinhua 321004, China}
\email{wangpei@zjnu.cn}

\begin{abstract}
We develop a Schr\"{o}dinger-picture formulation for a scalar quantum
field driven by a Lorentz-invariant white-noise field. The quantum
state of the system is described by a stochastic wave functional that
evolves according to a stochastic Schr\"{o}dinger equation. We show
that the Gaussian structure of the wave functional is preserved under
the stochastic evolution, allowing the dynamics to be reduced to a set
of equations for the corresponding kernel functions. These kernel
equations are derived and solved exactly, yielding an explicit
time-dependent expression for the wave functional.
The exact solution enables a direct analysis of the statistical
properties of the quantum field in the space of field configurations.
In particular, we show that the expectation value of the field
operator obeys the same stochastic equation as the classical field
obtained from the Euler-Lagrange equation of the
action. We further compute the energy density from the stochastic
wave functional and evaluate its ensemble average over noise
realizations. The resulting energy production rate coincides with
that obtained from the corresponding Lindblad equation.
This result indicates that the stochastic quantum state remains well
defined even though certain derived observables exhibit ultraviolet
divergences associated with the white-noise idealization.
\end{abstract}


\maketitle

\section{Introduction}
\label{sec:intro}

Stochastic formulations of quantum dynamics arise naturally in several areas 
of modern theoretical physics. In the study of open quantum systems, the 
evolution of a subsystem interacting with an environment is commonly 
described by quantum master equations such as the Lindblad equation. 
These equations admit stochastic unravelings in terms of Schr\"{o}dinger 
equations supplemented by random noise terms, providing a trajectory-level 
description of the dynamics~\cite{Gisin92,Percival98,Plenio98,Daley14,Castin95,Power96,Ates12,Hu13,Raghunandan18,Pokharel18,Weimer21,Weimer22}. Stochastic Schr\"{o}dinger equations also appear 
in proposals for spontaneous wave-function collapse~\cite{GRW,Diosi89,CSL,CSL2,Ghirardi90,Penrose96,Pearle99,Bassi05,Adler07,Adler08,Bassi13,Vinante16,Vinante17,Bahrami18,Tilloy19,Pontin19,Vinante20,Zheng20,Komori20,Donadi21,Gasbarri21,Carlesso22}, where stochastic 
modifications of quantum dynamics are introduced to address foundational 
questions concerning the emergence of random outcomes in measurements. 
These developments motivate the investigation of stochastic extensions of 
quantum theory at a more fundamental level, including their possible 
formulation within relativistic quantum field theory (QFT)~\cite{Ghirardi90,Pearle99,Myrvold17,Tumulka20,Jones20,Jones21,Wang22,Wang24}.

Relativistic extensions of stochastic quantum dynamics have recently attracted 
attention in the context of QFT~\cite{Wang22,Wang24}. In particular, a stochastic scalar field theory 
was proposed in Ref.~[\onlinecite{Wang22}], where a real scalar field is coupled to a 
Lorentz-invariant white-noise source and the dynamics is formulated in 
Hilbert space. This framework provides a relativistic generalization of 
stochastic Schr\"{o}dinger dynamics and allows one to investigate stochastic 
modifications of QFT while preserving Lorentz symmetry at the level of 
statistical laws.

However, the quantization of such stochastic field theories raises nontrivial 
difficulties. In particular, when canonical quantization and the conventional 
scattering-matrix framework are applied, ultraviolet divergences appear in 
quantities such as the scattering matrix~\cite{Wang22} and the energy production rate. 
These divergences indicate that the standard $S$-matrix formulation, which 
assumes asymptotic evolution over infinite time intervals, may not be the most 
appropriate framework for quantizing this class of stochastic theories. This 
observation motivates the search for alternative formulations of the quantum 
dynamics that remain well defined while preserving the essential stochastic 
structure of the theory.

In this paper we develop an alternative formulation based on the 
Schr\"{o}dinger picture of QFT. In this approach, the wave 
functional is expressed as a functional of the scalar field, while the canonical 
momentum operator is represented by functional derivatives with respect to 
the field. We study the stochastic time evolution of the wave functional over a 
finite time interval, starting from an initial state that is assumed to have a 
Gaussian form with time-dependent kernel functions. We show that this 
Gaussian functional form is preserved under the stochastic evolution, so that 
the dynamics reduces to a closed set of equations for the kernel functions. 
Solving these equations yields the complete time evolution of the wave 
functional. Within this framework, the quantum expectation value of the scalar 
field is found to satisfy exactly the classical stochastic field equation, namely 
the Euler-Lagrange equation derived from the classical theory, thereby 
establishing the classical-quantum correspondence of the present 
Schr\"{o}dinger-picture formulation. 

We also evaluate the expectation value of the energy density with respect to 
the Gaussian wave functional. The resulting energy production rate exhibits 
an ultraviolet divergence originating from the white-noise idealization, in 
agreement with previous density-matrix analyses. Moreover, the energy 
density itself becomes ultraviolet divergent for $t>0$, while remaining finite 
at the initial time. The divergences of the energy density and its production 
rate are mutually consistent within the present formulation.

Multi-component scalar fields of the type analyzed in this work arise in various 
contexts of relativistic QFT. In particular, the Higgs sector of 
the Standard Model can be decomposed into independent real scalar 
components. The present Schr\"{o}dinger-picture formulation therefore provides 
a mathematically controlled framework that may be relevant for stochastic 
extensions of such scalar sectors. Further exploration of these possibilities is left for future work.

The remainder of this paper is organized as follows. 
In Sec.~\ref{sec:review} we briefly review the Schr\"{o}dinger-picture formulation
of a free scalar quantum field. In Sec.~\ref{sec:White} we introduce the 
stochastic extension by coupling the field to a Lorentz-invariant 
white-noise source and derive the resulting stochastic evolution 
equation for the wave functional. In Sec.~\ref{sec:wf} we propose a 
Gaussian wave-functional ansatz, obtain the closed evolution equations 
for the corresponding kernel functions, and solve them exactly. 
In Sec.~\ref{sec:cq} we analyze the expectation value of the quantum 
field and show that it satisfies the classical Euler-Lagrange equation. 
In Sec.~\ref{sec:ed} we study the energetic properties of the system, 
computing the energy density from the stochastic wave functional and 
comparing the resulting energy production rate with that obtained from 
the Lindblad equation. Finally, Sec.~\ref{sec:con} contains our 
conclusions and discussion.

\section{Schr\"{o}dinger-picture formulation of scalar quantum field theory}
\label{sec:review}

In this section we briefly review the Schr\"{o}dinger-picture formulation of scalar 
QFT. In this formulation, quantum states are represented by wave functionals of 
the field configuration, and their time evolution is governed by a functional 
Schr\"{o}dinger equation. This formulation provides a direct description of 
real-time dynamics over finite time intervals, in contrast to approaches based 
on asymptotic scattering states. For a real scalar field $\phi(\mathbf{x})$, the 
wave functional $\Psi[\phi,t]$ depends on the spatial field configuration 
$\phi(\mathbf{x})$ at time $t$, while operators act on the wave functional through 
multiplication by the field or through functional derivatives with respect to 
the field.

The Lorentz-invariant action of the free scalar field theory is
\begin{equation}
S_0 = \int d^4 x \left( -\frac{1}{2} \partial_\mu \phi \partial^\mu \phi 
-\frac{1}{2} m^2 \phi^2 \right),
\end{equation}
where $x = (t,\mathbf{x})$ denotes the spacetime coordinates and $m$ is the 
mass of the scalar field. Throughout this paper we assume a flat Minkowski 
spacetime with metric signature $(-,+,+,+)$, and we set $\hbar=c=1$.

Applying canonical quantization, the Legendre transformation of the action 
leads to the Hamiltonian
\begin{equation}
\hat{H}_0 =
\int d^3 \mathbf{x} \left[
\frac{1}{2} \hat{\pi}^2(\mathbf{x})
+ \frac{1}{2} (\nabla \hat{\phi}(\mathbf{x}))^2
+ \frac{1}{2} m^2 \hat{\phi}^2(\mathbf{x})
\right],
\end{equation}
where $\hat{\phi}(\mathbf{x})$ and $\hat{\pi}(\mathbf{x})$ are the canonical 
field and momentum operators, respectively. They satisfy the equal-time commutation relation
$\left[\hat{\phi}(\mathbf{x}),\hat{\pi}(\mathbf{x}') \right]
= i \delta^3(\mathbf{x}-\mathbf{x}')$.

In the Schr\"{o}dinger-picture formulation, the canonical momentum operator is 
represented as a functional derivative,
\begin{equation}
\hat{\pi}(\mathbf{x}) =
- i \frac{\delta}{\delta \phi(\mathbf{x})},
\end{equation}
so that the Hamiltonian operator becomes
\begin{equation}
\hat{H}_0 =
\int d^3 \mathbf{x} \left[
- \frac{1}{2} \frac{\delta^2}{\delta \phi(\mathbf{x})^2}
+ \frac{1}{2} (\nabla \phi(\mathbf{x}))^2
+ \frac{1}{2} m^2 \phi^2(\mathbf{x})
\right].
\end{equation}
The Hamiltonian $\hat{H}_0$ is self-adjoint and generates deterministic and 
unitary evolution of the quantum state.

For later convenience, it is useful to work in the momentum-space basis. 
In this basis the field configuration can be expressed through its 
Fourier components,
\begin{equation}
\phi(\mathbf{p}) =
\frac{1}{(2\pi)^3}
\int d^3 \mathbf{x}\,
\phi(\mathbf{x}) e^{-i\mathbf{p}\cdot\mathbf{x}} .
\end{equation}
The reality condition of the field $\phi(\mathbf{x})$ implies
$\phi(\mathbf{p}) = \phi^*(-\mathbf{p})$. Independent variables in momentum 
space may therefore be chosen as the real and imaginary parts of the complex 
quantity $\phi(\mathbf{p})$, denoted by $\phi_R(\mathbf{p})$ and 
$\phi_I(\mathbf{p})$, with the momentum domain restricted to half of 
momentum space. For example, one may take $\mathbf{p}\in\Pi^+$, where 
$\Pi^+$ denotes the region with $p^3>0$.

In momentum space the Hamiltonian operator can then be written as
\begin{equation}\label{eq:S:H}
\begin{split}
\hat{H}_0
= & \ (2\pi)^3 \int_{\Pi^+} d^3 \mathbf{p}\,
E_{\mathbf{p}}^2 |\phi(\mathbf{p})|^2 \\
& - \frac{1}{4(2\pi)^3}
\int_{\Pi^+} d^3 \mathbf{p}
\left(
\frac{\delta^2}{\delta \phi_R(\mathbf{p})^2}
+ \frac{\delta^2}{\delta \phi_I(\mathbf{p})^2}
\right),
\end{split}
\end{equation}
where $E_{\mathbf{p}} = \sqrt{\mathbf{p}^2 + m^2}$ is the relativistic dispersion relation.

The wave functional then satisfies the functional Schr\"{o}dinger equation
\begin{equation}\label{eq:S:S}
i \frac{\partial}{\partial t} \Psi[\phi,t]
= \hat{H}_0 \Psi[\phi,t],
\end{equation}
which governs the deterministic evolution of the quantum state over an 
arbitrary finite time interval.

\section{Inclusion of White Noise}
\label{sec:White}

To introduce stochastic effects while preserving Lorentz invariance, it is 
convenient to formulate the theory at the level of the action. Following 
Ref.~[\onlinecite{Wang22}], a Lorentz-invariant stochastic field theory can be obtained by 
adding to the free-field action $S_0$ a coupling between the scalar field 
$\phi(x)$ and a white-noise field $dW(x)$. The quantity $dW(x)$ is defined as a 
set of independent Gaussian random variables associated with each spacetime 
cell of volume $d^4x$. Its mean value vanishes and its variance is given by 
$d^4x$.

With this definition, the statistical properties of $dW(x)$ are invariant under 
spacetime translations and Lorentz transformations. In this sense, $dW(x)$ may 
be regarded as a scalar field in the statistical sense. A more detailed 
discussion of this construction and the associated statistical Lorentz symmetry 
can be found in Ref.~[\onlinecite{Wang22}].

The complete action then takes the form
\begin{equation}\label{eq:W:S}
S =
\int d^4 x \left(
-\frac{1}{2} \partial_\mu \phi \partial^\mu \phi
-\frac{1}{2} m^2 \phi^2
\right)
+ \lambda \int dW(x)\,\phi(x),
\end{equation}
where $\lambda$ denotes the coupling constant between the scalar field and the 
noise. The notation $\int dW(x)$ represents a stochastic integral, which may be 
defined as the limit
\be
\int dW(x) = \lim_{d^4x\to0}\sum_x dW(x),
\ee
where the sum runs over spacetime cells.

The canonical quantization of such a stochastic action has been discussed in 
detail in Ref.~[\onlinecite{Wang22}]. As in the conventional case, one performs a Legendre 
transformation of the action. However, a technical subtlety arises due to the 
presence of the noise increment $dW(x)$. In particular, the ratio 
$dW(x)/d^4x$ is not well defined in the limit $d^4x\to0$. For this reason it is 
more convenient to formulate the dynamics in terms of an \emph{infinitesimal 
Hamiltonian integral} rather than a Hamiltonian itself.

The infinitesimal Hamiltonian integral can be viewed as the Hamiltonian 
multiplied by an infinitesimal time interval $dt$, and it generates the 
evolution of the quantum state over this interval. For the action 
\eqref{eq:W:S}, the corresponding Hamiltonian integral is
\begin{equation}
d\hat{H}_t
=
dt\,\hat{H}_0
-
\lambda
\int_{\mathbf{x}}
dW(t,\mathbf{x})\,{\phi}(\mathbf{x}),
\end{equation}
where $\hat{H}_0$ is the free-field Hamiltonian defined in 
Eq.~\eqref{eq:S:H}. Here the notation
$\int_{\mathbf{x}} dW(t,\mathbf{x})$
denotes a stochastic integral over the spatial coordinate only, which may be 
defined as
\be
\int_{\mathbf{x}} dW(t,\mathbf{x})
=
\lim_{dt,\,d^3\mathbf{x}\to0}
\sum_{\mathbf{x}} dW(t,\mathbf{x}).
\ee
The unitary evolution operator over the infinitesimal time interval from $t$ 
to $t+dt$ is therefore
\begin{equation}
\begin{split}
\hat{U}
&=
\exp\!\left\{-i\,d\hat{H}_t\right\} \\
&=
\exp\!\left\{
-i\,dt\,\hat{H}_0
+i\lambda
\int_{\mathbf{x}} dW(t,\mathbf{x})\,{\phi}(\mathbf{x})
\right\}.
\end{split}
\end{equation}

Given the wave functional at time $t$, the wave functional at time $t+dt$ is
\begin{equation}\label{eq:W:Ps}
\Psi[\phi,t+dt] = \hat{U}\,\Psi[\phi,t].
\end{equation}
Once the initial wave functional is specified (which we take at $t_0=0$ without 
loss of generality), the wave functional at any later time can be obtained by 
iteratively applying Eq.~\eqref{eq:W:Ps}. An equivalent functional derivative 
equation for $\Psi$, analogous to Eq.~\eqref{eq:S:S}, can also be derived 
straightforwardly. However, the evolution equation~\eqref{eq:W:Ps} is already 
sufficient for deriving the dynamics of the kernel functions in the wave 
functional ansatz considered below.

After the inclusion of the noise term, the deterministic functional 
Schr\"{o}dinger equation of the free theory is replaced by a stochastic 
evolution equation for the wave functional. Consequently, the evolution of 
$\Psi[\phi,t]$ becomes a stochastic process. In this formulation the wave 
functional remains the primary dynamical object, and for a given initial state 
the wave functional at any finite time $t$ is well defined. At the same time, 
observables constructed from composite operators may exhibit additional 
ultraviolet sensitivities associated with the idealized white-noise 
correlation.

\section{Wave Functional Ansatz and Kernel Equations}
\label{sec:wf}

In this section we analyze the stochastic evolution of the wave functional.
Because the Hamiltonian integral $d\hat{H}_t$
is at most quadratic in the field operator ${\phi}$, a Gaussian
wave functional remains Gaussian under the stochastic evolution.
It is therefore natural to adopt a Gaussian ansatz in which the kernel
functions evolve in time.

By inserting this ansatz into the evolution equation
\eqref{eq:W:Ps}, one can derive a closed set of equations for the
kernel functions. Remarkably, these equations admit exact solutions,
allowing the wave functional to be obtained at arbitrary time.

\subsection{Gaussian ansatz for the wave functional}

In general, a Gaussian wave functional can be written in the form
\begin{equation}
\begin{split}
\Psi[\phi,t] = \exp \Bigg\{ 
& -\frac{1}{2} \int d^3 \mathbf{x} \, d^3\mathbf{x}' \,
V(t, \mathbf{x}, \mathbf{x}') 
\phi(\mathbf{x}) \phi(\mathbf{x}')  \\
& + \int d^3 \mathbf{x} \, \mu(t,\mathbf{x}) \phi(\mathbf{x}) 
+ \mathcal{N}(t) \Bigg\},
\end{split}
\end{equation}
where $V(t, \mathbf{x}, \mathbf{x}')$ and $\mu(t,\mathbf{x})$ are kernel
functions to be determined, and $\mathcal{N}(t)$ is a normalization factor
whose explicit value is not relevant for the physical results considered here.

The quadratic kernel $V(t,\mathbf{x},\mathbf{x}')$ determines the
fluctuation properties of the field configuration $\phi(\mathbf{x})$.
In particular, the diagonal part (or equivalently the inverse kernel)
controls the local variance of the field, while the off-diagonal part
encodes correlations between field configurations at different spatial
positions. On the other hand, the linear kernel $\mu(t,\mathbf{x})$
determines the mean value of the field configuration.

According to the structure of the quadratic term in the wave functional,
one naturally requires
$V(t,\mathbf{x},\mathbf{x}') = V(t,\mathbf{x}',\mathbf{x})$,
i.e., the kernel must be symmetric under exchange of its arguments.
However, $V(t,\mathbf{x},\mathbf{x}')$ may in general be a complex-valued
function, so is $\mu(t,\mathbf{x})$.

As will be shown below, the white-noise field does not influence the
dynamical equation of $V(t, \mathbf{x}, \mathbf{x}')$, but only affects
the evolution of $\mu(t,\mathbf{x})$. Since the dynamics of
$V(t, \mathbf{x}, \mathbf{x}')$ is therefore completely determined by
the free action $S_0$, which preserves spatial translation symmetry,
it is natural to assume that $V(t, \mathbf{x}, \mathbf{x}')$ is invariant
under spatial translations. This assumption greatly simplifies the
structure of the kernel and the corresponding dynamical equations.

Under this assumption it is convenient to rewrite the wave functional
in momentum space, where the spatial translation symmetry becomes
manifest. Using the Fourier representation of the field,
the Gaussian ansatz can be written as
\begin{equation}\label{eq:Wa:Pp}
\begin{split}
\Psi[\phi,t] = \exp \Bigg\{ 
& -\frac{(2\pi)^3}{2} \int d^3 \mathbf{p} \,
V(t, \mathbf{p}) \left| \phi(\mathbf{p}) \right|^2  \\
& + (2\pi)^3 \int d^3 \mathbf{p} \,
\mu(t,\mathbf{p}) \phi(\mathbf{p})
+ \mathcal{N}(t) \Bigg\},
\end{split}
\end{equation}
where $V(t, \mathbf{p})$ and $\mu(t,\mathbf{p})$ denote the kernel
functions in momentum space. They are related to the real-space kernels
through the Fourier transformations
\be
V(t, \mathbf{x}, \mathbf{x}')
=
\frac{1}{(2\pi)^3}
\int d^3\mathbf{p}\,
V(t, \mathbf{p})
e^{i \mathbf{p}\cdot (\mathbf{x}-\mathbf{x}')},
\ee
and
\be
\mu(t, \mathbf{x})
=
\int d^3 \mathbf{p} \,
\mu (t,\mathbf{p})  
e^{-i \mathbf{p}\cdot \mathbf{x}}.
\ee
The symmetry of the quadratic kernel in real space
implies the corresponding constraint in momentum space, $V(t,\mathbf{p}) = V(t,-\mathbf{p})$.

In what follows, Eq.~\eqref{eq:Wa:Pp} will be taken as the working ansatz
for the wave functional. By substituting this form into the evolution
equation of the wave functional, we will derive the dynamical equations
for the kernel functions $V(t, \mathbf{p})$ and $\mu(t,\mathbf{p})$.

\subsection{Evolution equations for the kernels}

We consider the evolution over an infinitesimal time step from $t$ to $t+dt$,
which is governed by Eq.~\eqref{eq:W:Ps}. The infinitesimal evolution operator
can be factorized as
\begin{equation}\label{eq:Ev:U}
\hat{U}
=
\exp\!\left\{
-i\,dt\,\hat{H}_0
\right\}
\exp\!\left\{
i\lambda
\int_{\mathbf{x}} dW(t,\mathbf{x})\,{\phi}(\mathbf{x})
\right\}.
\end{equation}
This factorization follows from the Baker-Campbell-Hausdorff formula.
The neglected terms are of order $dt\, dW(t,\mathbf{x})$, 
which can be ignored in the limit $dt\to 0$.

The operator $\hat{U}$ acts on the wave functional through two successive
operations. The right-most exponential is a functional of $\phi$ and
thus acts simply by multiplication. Its exponent can be expressed in
momentum space as
\begin{equation}
\int_{\mathbf{x}} dW(t,\mathbf{x})\,{\phi}(\mathbf{x})
=
(2\pi)^3
\int d^3\mathbf{p} \,
\phi(\mathbf{p})\, dW(t,\mathbf{p}),
\end{equation}
where the momentum-space noise field is defined by the Fourier transform
\begin{equation}\label{eq:Ev:dWp}
dW(t,\mathbf{p})
=
\frac{1}{(2\pi)^3}
\int_{\mathbf{x}} dW(t,\mathbf{x})
\,e^{i\mathbf{p}\cdot\mathbf{x}} .
\end{equation}

The statistical properties of $dW(t,\mathbf{p})$ follow directly from
those of $dW(t,\mathbf{x})$. Since $dW(t,\mathbf{x})$ are independent
Gaussian random variables with variance $dt\,d^3\mathbf{x}$, the
Fourier transformation represents a linear transformation of a vector
of independent Gaussian variables. Consequently, the real and imaginary
parts of $dW(t,\mathbf{p})$ remain independent Gaussian variables.

For clarity, it is convenient to temporarily discretize space into cells
of volume $d^3\mathbf{x}$ and assume a finite spatial volume $\Omega$.
The limit $\Omega\to\infty$ can be taken after the calculation.
Under this discretization the momentum space is also discrete,
with spacing $d^3\mathbf{p}=(2\pi)^3/\Omega$.

Denoting the real and imaginary parts of $dW(t,\mathbf{p})$ by
$dW_R(t,\mathbf{p})$ and $dW_I(t,\mathbf{p})$, respectively, one finds
that the set
\be
\left(
dW_R(t,\mathbf{p}_1),
dW_I(t,\mathbf{p}_1),
dW_R(t,\mathbf{p}_2),
dW_I(t,\mathbf{p}_2),
\ldots
\right)
\ee
forms a vector obtained from the vector
\be
\left(
dW(t,\mathbf{x}_1),
dW(t,\mathbf{x}_2),
\ldots
\right)
\ee
by an orthogonal linear transformation. Here the momentum variable
$\mathbf{p}$ runs over the half momentum space $\Pi^+$.
Therefore these variables are independent Gaussian random variables
with zero mean and variance
\begin{equation}
D = \frac{\Omega\,dt}{2(2\pi)^6}.
\end{equation}

Having clarified the properties of the noise term in
Eq.~\eqref{eq:Ev:U}, we now evaluate the action of the left-most
exponential. Expanding to first order in $dt$ gives
\begin{equation}
\exp\left\{-i dt \hat{H}_0\right\}
=
1 - i dt\,\hat{H}_0 + \mathcal{O}(dt^2),
\end{equation}
and the higher-order terms can be neglected.

Since $\hat{H}_0$ is expressed in terms of functional derivatives
with respect to $\phi(\mathbf{p})$, its action on the Gaussian
ansatz \eqref{eq:Wa:Pp} can be computed straightforwardly.
After a lengthy but straightforward calculation, one obtains
the wave functional at time $t+dt$. By comparing the resulting
expression with the Gaussian ansatz evaluated at $t+dt$, the
evolution equations for the kernels are obtained:
\begin{equation}\label{eq:Ev:dVmu}
\begin{split}
\frac{\partial}{\partial t} V(t,\mathbf{p})
&=
-i V^2(t,\mathbf{p})
+ i E^2_\mathbf{p},
\\
d_t \mu(t,\mathbf{p})
&=
-i dt \,\mu(t,\mathbf{p}) V(t,\mathbf{p})
+ i\lambda\, dW(t,\mathbf{p}),
\end{split}
\end{equation}
where $d_t\mu(t,\mathbf{p})=\mu(t+dt,\mathbf{p})-\mu(t,\mathbf{p})$
denotes the time increment.

The second equation is a stochastic differential equation.
To avoid the appearance of the ill-defined quantity
$dW(t,\mathbf{p})/dt$, it is written in the increment form,
as is standard in stochastic calculus.

Equation~\eqref{eq:Ev:dVmu} constitutes the dynamical equations
for the kernel functions. In particular, the equation for
$V(t,\mathbf{p})$ is independent of the noise field, confirming
the observation made in the previous subsection. Moreover,
in momentum space the equations for different momenta are
completely decoupled, which makes an analytical solution possible.

\subsection{Exact solution of the kernel equations}

The kernel equations~\eqref{eq:Ev:dVmu} can be solved exactly once the
initial conditions are specified, i.e., once the kernel functions at
$t=0$ are given. We denote the initial kernels by
$V_0(\mathbf{p}) = V(0,\mathbf{p})$ and
$\mu_0(\mathbf{p}) = \mu(0,\mathbf{p})$.

The equation for $V(t,\mathbf{p})$ is a Riccati equation. Its solution
can be written as
\begin{equation}\label{eq:ES:Vt}
V(t,\mathbf{p})
=
\frac{
V_0(\mathbf{p}) \cos(tE_{\mathbf{p}})
+ i E_{\mathbf{p}} \sin(tE_{\mathbf{p}})
}{
\cos(tE_{\mathbf{p}})
+ i \frac{V_0(\mathbf{p})}{E_{\mathbf{p}}}
\sin(tE_{\mathbf{p}})
}.
\end{equation}
The quadratic kernel is therefore a deterministic function of time
and is independent of the noise field. Equation~\eqref{eq:ES:Vt}
coincides with the kernel evolution of the free Schr\"{o}dinger-picture
scalar field, confirming that the stochastic noise affects only the mean
field but not the fluctuation kernel.

According to the Gaussian ansatz~\eqref{eq:Wa:Pp}, the inverse of
$V(t,\mathbf{p})$ characterizes the variance of the field component
$\phi(\mathbf{p})$. In this sense, the limits $V(t,\mathbf{p})\to0$
and $V(t,\mathbf{p})\to\infty$ correspond to an infinitely broad
distribution of $\phi(\mathbf{p})$ (complete randomness) and a
vanishing width (deterministic field value), respectively.

For a massive field with $m>0$, one has $E_{\mathbf{p}}>0$ for all
$\mathbf{p}$. Provided that $V_0(\mathbf{p})\neq0$, the solution
$V(t,\mathbf{p})$ remains finite for arbitrary time. Moreover,
the quadratic kernel oscillates with period $2\pi/E_{\mathbf{p}}$.
At the special times
$t = n\pi/E_{\mathbf{p}}$ with $n=1,2,\ldots$, one finds
$V(t,\mathbf{p})=V_0(\mathbf{p})$, while at
$t=(n+\tfrac12)\pi/E_{\mathbf{p}}$ the kernel becomes
$V(t,\mathbf{p}) = E_{\mathbf{p}}^2/V_0(\mathbf{p})$.

It is instructive to consider several specific initial conditions.
If $V_0(\mathbf{p})=E_{\mathbf{p}}$, one obtains
$V(t,\mathbf{p})\equiv E_{\mathbf{p}}$, so that the quadratic kernel
remains constant in time. This corresponds to the stationary kernel
of the free-field vacuum.

If instead $V_0(\mathbf{p})=0$, corresponding to an initially
divergent variance of $\phi(\mathbf{p})$, Eq.~\eqref{eq:ES:Vt}
gives $V(t,\mathbf{p}) = iE_{\mathbf{p}}\tan(tE_{\mathbf{p}})$,
which is purely imaginary and whose magnitude oscillates between
zero and infinity.

Conversely, if $V_0(\mathbf{p})\to\infty$, corresponding to a
deterministic initial field configuration, one obtains
$V(t,\mathbf{p}) = -iE_{\mathbf{p}}\cot(tE_{\mathbf{p}})$,
which also exhibits strong oscillations between zero and infinity
as time evolves.

We now turn to the linear kernel $\mu(t,\mathbf{p})$, which determines
the mean value of the field configuration $\phi(\mathbf{p})$.
Unlike $V(t,\mathbf{p})$, the evolution of $\mu(t,\mathbf{p})$ is
directly driven by the noise. The solution of its stochastic
differential equation can be written in the general form
\begin{equation}\label{eq:ES:mu}
\mu(t,\mathbf{p})
=
e^{-i\int_0^t d\tau\, V(\tau,\mathbf{p})}\,
\mu_0(\mathbf{p})
+
i\lambda
\int_0^t dW(\tau,\mathbf{p})
\,
e^{-i\int_\tau^t ds\, V(s,\mathbf{p})}.
\end{equation}

The first term represents the deterministic contribution arising
from the free evolution governed by $\hat{H}_0$, while the second
term describes the stochastic contribution induced by the noise
field.

A physically relevant choice of initial condition is
$\mu_0(\mathbf{p})\equiv0$, which corresponds to an initial field
distribution symmetric about $\phi=0$. In this case the deterministic
contribution vanishes and the linear kernel reduces to
\begin{equation}\label{eq:ES:m1}
\mu(t,\mathbf{p})
=
i\lambda
\int_0^t dW(\tau,\mathbf{p})
\,
e^{-i\int_\tau^t ds\, V(s,\mathbf{p})}.
\end{equation}

Equation~\eqref{eq:ES:m1} shows that $\mu(t,\mathbf{p})$ is a linear
combination of the Gaussian random variables $dW(\tau,\mathbf{p})$.
Therefore $\mu(t,\mathbf{p})$ itself follows a Gaussian distribution
with zero mean. Its variance can be obtained straightforwardly from
the noise correlation and will be analyzed below.

Physically, the white-noise field drives the expectation value of the
field configuration into a stochastic motion in field-configuration
space. A more detailed analysis of the statistical properties of
$\phi$ will be presented in the following sections.

\section{Classical-to-Quantum Correspondence}
\label{sec:cq}

In this section we establish a classical-to-quantum correspondence
within the Schr\"{o}dinger formulation developed above.
Specifically, we calculate the quantum expectation value of the field
operator and show that it satisfies exactly the same dynamical equation
as the classical field obtained from the Euler-Lagrange principle.

We first recall the classical equation of motion.
By applying the extremal condition $\delta S = 0$ to the random-valued
action, the classical field $\phi(t,\mathbf{x})$ satisfies
\begin{equation}
d^4 x\left(
-\partial_t^2 \phi(x)
+\nabla^2 \phi(x)
- m^2 \phi(x)
\right)
+ \lambda dW(x)
=0 .
\end{equation}

For later comparison it is convenient to rewrite this equation in
momentum space. Using the inverse Fourier transformation of
Eq.~\eqref{eq:Ev:dWp}, $dW(t,\mathbf{x})
= d^3\mathbf{x}
\int d^3\mathbf{p}\,
e^{-i\mathbf{p}\cdot\mathbf{x}}
dW(t,\mathbf{p})$,
together with the property
$dW(t,-\mathbf{p}) = dW^*(t,\mathbf{p})$,
the classical equation becomes
\be\label{eq:qc:dcp}
dt\left(
\partial_t^2 \phi(t,\mathbf{p})
+ E_{\mathbf{p}}^2 \phi(t,\mathbf{p})
\right) -
\lambda dW^*(t,\mathbf{p})
=0 .
\ee

We now turn to the quantum description.
The quantum expectation value of the field operator is defined as
\be\label{eq:qc:q}
\begin{split}
\phi^{(q)}(t,\mathbf{x})
&=
\langle \Psi_t | \hat{\phi}(\mathbf{x}) | \Psi_t \rangle  \\
&=
\int D\phi \,
|\Psi[\phi,t]|^2
\,\phi(\mathbf{x}),
\end{split}
\ee
where $|\Psi_t\rangle$ denotes the quantum state of the field at time
$t$. The expectation value can therefore be expressed as a functional
integral over field configurations with probability density
$P[\phi,t]=|\Psi[\phi,t]|^2$.
The quantity $\phi^{(q)}(t,\mathbf{x})$ thus plays the role of the
quantum counterpart of the classical field $\phi(t,\mathbf{x})$.

Since the wave functional has been expressed in momentum space,
it is convenient to evaluate the expectation value in that
representation, namely
$\phi^{(q)}(t,\mathbf{p})
= \langle \Psi_t | \hat{\phi}(\mathbf{p}) | \Psi_t \rangle$.
Using the Gaussian ansatz~\eqref{eq:Wa:Pp}, the probability density of
the field configuration takes the form
\begin{widetext}
\be\label{eq:qc:PP}
\begin{split}
P[\phi,t]
=&
\exp\Bigg\{
-2(2\pi)^3
\int_{\Pi^+} d^3\mathbf{p}\,
V_R(t,\mathbf{p})
\left(
\phi_R(\mathbf{p})
-
\frac{\mu_R(t,\mathbf{p})+\mu_R(t,-\mathbf{p})}
{2V_R(t,\mathbf{p})}
\right)^2
\\
&
-
2(2\pi)^3
\int_{\Pi^+} d^3\mathbf{p}\,
V_R(t,\mathbf{p})
\left(
\phi_I(\mathbf{p})
-
\frac{\mu_I(t,-\mathbf{p})-\mu_I(t,\mathbf{p})}
{2V_R(t,\mathbf{p})}
\right)^2
+
\text{const.}
\Bigg\},
\end{split}
\ee
\end{widetext}
where $V_R$ and $V_I$ denote the real and imaginary parts of the
quadratic kernel $V$, and $\mu_R$ and $\mu_I$ denote the real and
imaginary parts of $\mu$. Field-independent constants have been
omitted.

Since the probability distribution is Gaussian, the expectation value
of the field can be read off directly from the center of the
distribution. One finds
\be\label{eq:cq:phe}
\phi^{(q)}(t,\mathbf{p})
=
\frac{
\mu^*(t,\mathbf{p})+\mu(t,-\mathbf{p})
}{2V_R(t,\mathbf{p})}.
\ee
The quantum expectation value is therefore expressed in terms of the
kernel functions whose dynamical equations were derived in the
previous section.
We can now determine the time evolution of
$\phi^{(q)}(t,\mathbf{p})$ by using those kernel equations.

From the stochastic differential equation for $\mu$ in
Eq.~\eqref{eq:Ev:dVmu}, one obtains
\be
\begin{split}
d_t
\left[
\mu^*(t,\mathbf{p})
+
\mu(t,-\mathbf{p})
\right]
=
idt
\left[
\mu^*(t,\mathbf{p})V^*(t,\mathbf{p})
-
\mu(t,-\mathbf{p})V(t,\mathbf{p})
\right].
\end{split}
\ee
An important observation is that the noise terms cancel in this
combination. Consequently, the time derivative of this quantity is
well defined:
\be
\begin{split}
\partial_t
\left[
\mu^*(t,\mathbf{p})
+
\mu(t,-\mathbf{p})
\right]
=
\mu^*(t,\mathbf{p})V^*(t,\mathbf{p})
-
\mu(t,-\mathbf{p})V(t,\mathbf{p}).
\end{split}
\ee
This should be contrasted with the isolated quantity $\mu(t,\mathbf{p})$,
for which the partial derivative $\partial_t \mu$ is not well defined due to
the presence of the stochastic increment $dW$.

The calculation of the second-order partial derivative of
$\phi^{(q)}(t,\mathbf{p})$ then becomes straightforward.
After a direct but somewhat lengthy algebra, and by using the
dynamical equation for $V(t,\mathbf{p})$, we obtain
\be\label{eq:qc:dpp}
dt\left(
\partial_t^2 \phi^{(q)}(t,\mathbf{p})
+
E_{\mathbf{p}}^2
\phi^{(q)}(t,\mathbf{p})
\right)
-
\lambda dW^*(t,\mathbf{p})
=
0 .
\ee
Comparing Eq.~\eqref{eq:qc:dpp} with the classical equation
Eq.~\eqref{eq:qc:dcp}, we see that the quantum expectation value of the
field obeys exactly the same dynamical equation as the classical
field.

This result establishes a direct classical-to-quantum correspondence
in the present stochastic field theory.
Although the quantum state of the field evolves according to a
stochastic Schr\"{o}dinger equation in the space of wave
functionals, the expectation value of the field operator follows the
same stochastic Euler-Lagrange equation that governs the classical
field. In this sense, the classical stochastic dynamics emerges
naturally as the evolution equation for the quantum expectation
value of the field. This property resembles the Ehrenfest theorem in 
quantum mechanics, where expectation values obey classical equations of motion.

\section{Energy Density and Energy Production Rate}
\label{sec:ed}

We now turn to the energetic properties of the theory.
A characteristic feature of couplings to a white-noise
background is the continuous injection of energy into the quantum
system. In quantum field theory this mechanism typically leads to an
ultraviolet-divergent energy production rate, a phenomenon widely
discussed in the literature on stochastic and collapse-type dynamical
models~\cite{Ghirardi90,Bassi13}.

In this section we analyze the energy production rate within the
present formulation. We first derive the rate from the
Lindblad equation for the density matrix. Since the Lindblad equation
describes the ensemble-averaged evolution of the quantum state, the
result obtained in this way corresponds to the energy production rate
averaged over all realizations of the noise field. This derivation is
included here in order to make the discussion self-contained.

We then compute the expectation value of the Hamiltonian directly
from the stochastic wave functional obtained in Sec.~\ref{sec:wf}. In this
approach the energy expectation value depends on the particular
noise trajectory. By performing the stochastic average over the noise
realizations, we obtain the ensemble-averaged energy density and the
corresponding production rate. As will be shown below, the final
result coincides with that derived from the Lindblad equation,
providing a nontrivial consistency check of the wave-functional
formalism.

\subsection{Energy production rate from the Lindblad equation}

We first derive the energy production rate from the Lindblad equation
corresponding to the stochastic evolution~\eqref{eq:W:Ps}. The
Lindblad equation is obtained by averaging the stochastic evolution
over the realizations of the noise field $dW(t,\mathbf{x})$. The
resulting equation for the density matrix $\hat{\rho}$ was derived in
our previous work~\cite{Wang22} and reads
\begin{equation}\label{eq:ep:dr}
\frac{d \hat{\rho}}{dt}
=
-i \left[ \hat{H}_0, \hat{\rho}\right]
+
\lambda^2
\int d^3\mathbf{x}
\left(
\hat{\phi}(\mathbf{x})\hat{\rho}\hat{\phi}(\mathbf{x})
-
\frac{1}{2}
\left\{ \hat{\phi}^2(\mathbf{x}), \hat{\rho}\right\}_+
\right),
\end{equation}
where $\hat{\rho}$ denotes the density matrix of the quantum field.

The ensemble-averaged energy density is defined as
\begin{equation}\label{eq:ep:dE}
E(t)
=
\frac{1}{\Omega}
\mathrm{Tr}\!\left(\hat{\rho}\,\hat{H}_0\right),
\end{equation}
where $\Omega$ is the spatial volume. The corresponding energy
production rate is therefore
\begin{equation}
\frac{dE(t)}{dt}
=
\frac{1}{\Omega}
\frac{d}{dt}
\mathrm{Tr}\!\left(\hat{\rho}\,\hat{H}_0\right).
\end{equation}

To evaluate this quantity it is convenient to express the field
operators in terms of the bosonic creation and annihilation operators
$\hat{a}_{\mathbf{p}}^\dagger$ and $\hat{a}_{\mathbf{p}}$. In momentum
space the free Hamiltonian takes the standard form $\hat{H}_0=
\int d^3\mathbf{p}\,
E_{\mathbf{p}}
\hat{a}_{\mathbf{p}}^\dagger
\hat{a}_{\mathbf{p}} $.
Substituting the Lindblad equation~\eqref{eq:ep:dr} into the time
derivative of Eq.~\eqref{eq:ep:dE} and using the canonical commutation
relation $[\hat{a}_{\mathbf{p}},\hat{a}_{\mathbf{p}'}^\dagger]
=\delta^3(\mathbf{p}-\mathbf{p}')$,
one finds after straightforward algebra
\begin{equation}
\frac{dE(t)}{dt}
=
\frac{\lambda^2}{2}
\frac{1}{\Omega}
\int d^3\mathbf{p}\,
\delta^3(0).
\end{equation}

The factor $\delta^3(0)$ reflects the continuum of the
momentum states. To interpret it, it is convenient to temporarily
consider a finite spatial volume $\Omega$. In this case the momentum
space becomes discretized with cell volume
$d^3\mathbf{p}=(2\pi)^3/\Omega$, and the Dirac delta function is
replaced by the Kronecker delta. Consequently, $\delta^3(0)=1/d^3\mathbf{p}$,
which leads to
\begin{equation}\label{eq:ed:dedt}
\frac{dE(t)}{dt}
=
\frac{\lambda^2}{2(2\pi)^3}
\int d^3\mathbf{p}.
\end{equation}

At this point a genuine ultraviolet divergence appears. Since the
momentum integral extends over arbitrarily large momenta and no
natural cutoff exists within the present relativistic field theory,
the integral $\int d^3\mathbf{p}$ diverges. This
result reflects the well-known problem of divergent energy production
in quantum systems driven by ideal white noise.

Physically, the divergence originates from the flat spectrum of white
noise, which injects energy into all momentum modes with equal
strength. Because arbitrarily high-momentum modes are present in the
field theory, the total injected energy becomes unbounded.

It is worth emphasizing that in the Schr\"{o}dinger formulation
developed in the previous sections the stochastic wave functional
remains mathematically well defined for arbitrary time, despite the
presence of the noise driving. At first sight this may appear to be in
tension with the divergent energy production rate derived above.
In the next subsection we compute the energy density directly from the
wave functional and show that, after performing the stochastic
average, the same energy production rate is recovered. This
demonstrates the consistency between the Lindblad and
wave-functional approaches.

\subsection{Energy density from the stochastic wave functional}

We now compute the energy density directly from the stochastic wave
functional obtained in Sec.~\ref{sec:wf}. For a given realization of the noise,
the quantum expectation value of the Hamiltonian is
\begin{equation}
\mathcal{E}(t) = \bra{\Psi_t} \hat{H}_0 \ket{\Psi_t}.
\end{equation}

To evaluate this quantity we first consider the action of $\hat{H}_0$
on the wave functional. As shown in Eq.~\eqref{eq:S:H}, the Hamiltonian
operator can be written in terms of functional derivatives with
respect to the field configuration $\phi(\mathbf{p})$. Since the
wave functional $\Psi[\phi,t]$ has an exponential Gaussian form,
the action of the functional derivatives preserves this structure
and simply produces polynomial prefactors multiplying
$\Psi[\phi,t]$.

After a straightforward calculation one finds that the energy can be
decomposed into two contributions,
\begin{equation}
\mathcal{E}(t) = \mathcal{E}^{(0)}(t) + \mathcal{E}^{(1)}(t),
\end{equation}
where
\begin{equation}\label{eq:ed:E0}
\begin{split}
\mathcal{E}^{(0)}(t)
=
& \frac{\Omega}{2(2\pi)^3}
\int d^3\mathbf{p}\,V(t,\mathbf{p})
\\
& + \frac{(2\pi)^3}{2}
\int d^3\mathbf{p}\,
\left(E^2_{\mathbf{p}} -V^2(t,\mathbf{p})\right)
\bra{\Psi_t} \left| \phi(\mathbf{p})\right|^2 \ket{\Psi_t},
\end{split}
\end{equation}
represents the contribution independent of the stochastic driving,
while
\begin{equation}\label{eq:ed:E1}
\begin{split}
\mathcal{E}^{(1)}(t)
=
& - \frac{(2\pi)^3}{2}
\int d^3\mathbf{p}\,
\mu(t,\mathbf{p}) \mu(t,-\mathbf{p})
\\
& + {(2\pi)^3}
\int d^3\mathbf{p}\,
\mu(t,\mathbf{p}) V(t,\mathbf{p})
\bra{\Psi_t} \phi(\mathbf{p})\ket{\Psi_t},
\end{split}
\end{equation}
is the contribution induced by the stochastic noise through the
linear kernel $\mu(t,\mathbf{p})$.

The expectation values appearing above can be written as functional
integrals,
\begin{equation}\label{eq:ed:qep}
\begin{split}
\bra{\Psi_t} \left| \phi(\mathbf{p})\right|^2\ket{\Psi_t}
&=
\int D\phi\, |\Psi[\phi,t]|^2
\left| \phi(\mathbf{p})\right|^2 ,
\\
\bra{\Psi_t} \phi(\mathbf{p})\ket{\Psi_t}
&=
\int D\phi\, |\Psi[\phi,t]|^2
\phi(\mathbf{p}) .
\end{split}
\end{equation}
These quantities can be evaluated using the probability density
$P[\phi,t]=|\Psi[\phi,t]|^2$ derived in
Eq.~\eqref{eq:qc:PP}. Since the probability density is Gaussian,
the statistical properties of $\phi(\mathbf{p})$ can be read off
directly. The mean value $\bra{\Psi_t}\phi(\mathbf{p})\ket{\Psi_t}$ has already been obtained
in Sec.~\ref{sec:cq}, while the variance is given by
\begin{equation}\label{eq:ed:DP}
\begin{split}
\bra{\Psi_t} \phi_R^2(\mathbf{p}) \ket{\Psi_t}
-
\bra{\Psi_t} \phi_R(\mathbf{p}) \ket{\Psi_t}^2
&=
\frac{\Omega}{4(2\pi)^6 V_R(t,\mathbf{p})},
\\
\bra{\Psi_t} \phi_I^2(\mathbf{p}) \ket{\Psi_t}
-
\bra{\Psi_t} \phi_I(\mathbf{p}) \ket{\Psi_t}^2
&=
\frac{\Omega}{4(2\pi)^6 V_R(t,\mathbf{p})}.
\end{split}
\end{equation}

Using these relations together with the mean value given in
Eq.~\eqref{eq:cq:phe}, the expectation value
$\bra{\Psi_t}|\phi(\mathbf{p})|^2\ket{\Psi_t}$ can be obtained
explicitly. Substituting these results into
Eqs.~\eqref{eq:ed:E0} and~\eqref{eq:ed:E1}, the energy becomes
fully determined by the kernel functions $V(t,\mathbf{p})$ and
$\mu(t,\mathbf{p})$, whose time dependence has been solved in
Sec.~\ref{sec:wf}.

In particular, the noise-independent part reduces to
\begin{equation}
\mathcal{E}^{(0)}(t)
=
\frac{\Omega}{2(2\pi)^3}
\int d^3\mathbf{p}
\frac{E^2_{\mathbf{p}}+|V_0(\mathbf{p})|^2}
{2V_{0R}(\mathbf{p})},
\end{equation}
where $V_{0R}(\mathbf{p})$ denotes the real part of the initial
kernel $V_0(\mathbf{p})$. This contribution is independent of time
and therefore represents the initial energy of the quantum field.
The free part of the energy is thus unaffected by the stochastic
driving.

The noise-induced contribution $\mathcal{E}^{(1)}(t)$, however,
depends on the stochastic kernel $\mu(t,\mathbf{p})$ and therefore
varies from one noise realization to another. To compare with the
Lindblad description derived in the previous subsection, we must
compute its ensemble average over the noise distribution. Using the
explicit solution for $\mu(t,\mathbf{p})$ and the noise correlations,
one finds
\begin{equation}
\begin{split}
\overline{\mu(t,\mathbf{p}) \mu(t,-\mathbf{p})}
&=
-\lambda^2\frac{\Omega}{(2\pi)^6}
\int_0^t d\tau\,
e^{-2i\int_\tau^t ds\,V(s,\mathbf{p})},
\\
\overline{|\mu(t,\mathbf{p})|^2}
&=
\lambda^2\frac{\Omega}{(2\pi)^6}
\int_0^t d\tau\,
e^{2\int_\tau^t ds\,V_I(s,\mathbf{p})},
\end{split}
\end{equation}
where the overline denotes averaging over the noise ensemble.

Substituting these expressions into Eq.~\eqref{eq:ed:E1} yields the
ensemble-averaged noise contribution
\begin{equation}
\overline{\mathcal{E}^{(1)}(t)}
=
\frac{t\,\Omega\,\lambda^2}{2(2\pi)^3}
\int d^3\mathbf{p}.
\end{equation}
At the initial time $t=0$ this term vanishes, so that the total
energy coincides with the initial free-field energy
$\mathcal{E}^{(0)}$. For any $t>0$, however, the noise-induced
contribution grows linearly in time and exhibits an ultraviolet
divergence due to the unbounded momentum integral. The linear growth
reflects the continuous injection of energy into the quantum field
by the external noise.

Finally, the energy production rate follows from
\begin{equation}
\frac{dE(t)}{dt}
=
\frac{1}{\Omega}
\frac{d}{dt}
\left(
\mathcal{E}^{(0)}(t)
+
\overline{\mathcal{E}^{(1)}(t)}
\right).
\end{equation}
Since $\mathcal{E}^{(0)}$ is time independent, this yields $\frac{dE(t)}{dt}=
\frac{\lambda^2}{2(2\pi)^3}
\int d^3\mathbf{p}$,
which coincides exactly with the result obtained from the Lindblad
equation in the previous subsection.


The above calculation demonstrates that the stochastic wave-functional
formalism reproduces the same energy production rate as the Lindblad
density-matrix approach. Although the wave functional itself remains
well defined for each individual realization of the noise, the
ensemble-averaged energy diverges because the ideal white-noise
driving continuously injects energy into field modes of arbitrarily
high momentum. The ultraviolet divergence therefore originates from
the white-noise approximation rather than from any inconsistency of
the Schr\"{o}dinger formulation.

\section{Conclusion and Discussion}
\label{sec:con}

Quantum field theories driven by stochastic white noise have long been
known to exhibit pathological features, most notably the appearance of
ultraviolet-divergent energy production rates and related difficulties
in scattering descriptions. For this reason such models are often
regarded as physically problematic. In the present work we approached
this issue from a different perspective by developing a
Schr\"{o}dinger-picture formulation of stochastic quantum field
dynamics, in which the fundamental object is the stochastic wave
functional of the field configuration. This formulation provides a
trajectory-level description of the quantum dynamics and allows one to
analyze directly how the quantum state evolves under stochastic
driving.

For the case of a free scalar field subject to white-noise driving, we
showed that the stochastic evolution preserves the Gaussian structure
of the wave functional. This allows the dynamics to be reduced to a
set of equations for the kernel functions characterizing the Gaussian
state. These kernel equations were derived and solved exactly. The
quadratic kernel evolves deterministically and coincides with the
kernel of the free-field Schr\"{o}dinger evolution, while the linear
kernel encodes the stochastic influence of the noise. As a result, the
wave functional can be written explicitly at arbitrary time.

The exact solution makes it possible to analyze the statistical
properties of the quantum field directly in the space of field
configurations. In particular, we demonstrated a
classical-to-quantum correspondence within the Schr\"{o}dinger
framework: the expectation value of the field operator obeys precisely
the same stochastic equation as the classical field obtained from the
Euler-Lagrange equation of the action.

We also investigated the energetic properties of the system. The
energy density was computed directly from the stochastic wave
functional, first for individual noise trajectories and then after
averaging over the noise ensemble. The resulting energy production
rate agrees exactly with that obtained from the density-matrix
description, providing a nontrivial consistency check of the
Schr\"{o}dinger wave-functional formulation.

An important observation emerging from our analysis is that the
stochastic wave functional itself remains mathematically well defined
for all times, despite the presence of divergent energy production.
The divergence arises from the ideal white-noise spectrum, which
injects energy uniformly into field modes of arbitrarily high
momentum. From the viewpoint of the present formulation, the
ultraviolet divergence therefore reflects a limitation of the
white-noise idealization rather than an inconsistency of the quantum
state description. This suggests that stochastic quantum field
theories driven by white noise can still admit a consistent
microscopic description at the level of the quantum state, even though
certain derived observables become ultraviolet divergent.

More generally, the Schr\"{o}dinger wave-functional perspective
provides a natural framework for describing stochastic quantum field
dynamics at the level of individual trajectories. Because it gives
direct access to the statistical properties of field configurations,
this approach may be useful for studying more general stochastic
quantum field theories, including interacting fields and
nonequilibrium quantum systems. These directions will be explored in
future work.


\begin{thebibliography}{35}
\expandafter\ifx\csname natexlab\endcsname\relax\def\natexlab#1{#1}\fi
\expandafter\ifx\csname bibnamefont\endcsname\relax
  \def\bibnamefont#1{#1}\fi
\expandafter\ifx\csname bibfnamefont\endcsname\relax
  \def\bibfnamefont#1{#1}\fi
\expandafter\ifx\csname citenamefont\endcsname\relax
  \def\citenamefont#1{#1}\fi
\expandafter\ifx\csname url\endcsname\relax
  \def\url#1{\texttt{#1}}\fi
\expandafter\ifx\csname urlprefix\endcsname\relax\def\urlprefix{URL }\fi
\providecommand{\bibinfo}[2]{#2}
\providecommand{\eprint}[2][]{\url{#2}}

\bibitem{Gisin92} N. Gisin and I. C. Percival, {\it The quantum-state diffusion model applied to open systems}, J. Phys. A: Math. Gen. {\bf 25}, 5677 (1992).
\bibitem{Percival98} I. Percival, {\it Quantum State Diffusion} (Cambridge University Press, Cambridge, 1998).
\bibitem{Plenio98} M. B. Plenio and P. L. Knight, {\it The quantum-jump approach to dissipative dynamics in quantum optics}, Rev. Mod. Phys. {\bf 70}, 101 (1998).
\bibitem{Daley14} A. J. Daley, {\it Quantum trajectories and open many-body quantum systems}, Adv. Phys. {\bf 63}, 77 (2014).

\bibitem{Castin95} Y. Castin and K. M{\o}lmer, {\it Monte Carlo Wave-Function Analysis of 3D Optical Molasses}, Phys. Rev. Lett. {\bf 74}, 3772 (1995).
\bibitem{Power96} W. L. Power and P. L. Knight, {\it Stochastic simulations of the quantum Zeno effect}, Phys. Rev. A {\bf 53}, 1052 (1996).
\bibitem{Ates12} C. Ates, B. Olmos, J. P. Garrahan, and I. Lesanovsky, {\it Dynamical phases and intermittency of the dissipative quantum Ising model}, Phys. Rev. A
{\bf 85}, 043620 (2012).
\bibitem{Hu13} A. Hu, T. E. Lee, and C. W. Clark, {\it Spatial correlations of one-dimensional driven-dissipative systems of Rydberg atoms}, Phys. Rev. A {\bf 88}, 053627 (2013).
\bibitem{Raghunandan18} M. Raghunandan, J. Wrachtrup, and H. Weimer, {\it High-Density Quantum Sensing with Dissipative First Order Transitions}, Phys. Rev. Lett.
{\bf 120}, 150501 (2018).
\bibitem{Pokharel18} B. Pokharel, M. Z. Misplon, W. Lynn, P. Duggins, K. Hallman,
D. Anderson, A. Kapulkin, and A. K. Pattanayak, {\it Chaos and dynamical complexity in the quantum to classical transition}, Sci. Rep. {\bf 8}, 2108 (2018).
\bibitem{Weimer21} H. Weimer, A. Kshetrimayum, and R. Or\'{u}s, {\it Simulation methods for open quantum many-body systems}, Rev. Mod. Phys. {\bf 93}, 015008 (2021).
\bibitem{Weimer22} V. P. Singh and H. Weimer, {\it Driven-Dissipative Criticality within the Discrete Truncated Wigner Approximation}, Phys. Rev. Lett. {\bf 128}, 200602 (2022).

\bibitem{GRW} G. C. Ghirardi, A. Rimini, and T. Weber, {\it Unified dynamics for microscopic and macroscopic systems}, Phys. Rev. D {\bf 34}, 470 (1986).
\bibitem{Diosi89} L. Di\'{o}si, {\it Models for universal reduction of macroscopic quantum fluctuations}, Phys. Rev. A {\bf 40}, 1165 (1989).
\bibitem{CSL} P. Pearle, {\it Combining stochastic dynamical state-vector reduction with spontaneous localization}, Phys. Rev. A {\bf 39}, 2277 (1989).
\bibitem{CSL2} G. C. Ghirardi, P. Pearle, and A. Rimini, {\it Markov processes in Hilbert space and continuous spontaneous localization of systems of identical particles}, Phys. Rev. A {\bf 42}, 78 (1990).
\bibitem{Ghirardi90} G. C. Ghirardi, R. Grassi, and P. Pearle, {\it Relativistic dynamical reduction models: General framework and examples}, Foundations of Physics {\bf 20}, 1271 (1990).
\bibitem{Penrose96} R. Penrose, {\it On Gravity's role in Quantum State Reduction}, Gen. Relativ. Gravit. {\bf 28}, 581 (1996).
\bibitem{Pearle99} P. Pearle, {\it Relativistic collapse model with tachyonic features}, Phys. Rev. A {\bf 59}, 80 (1999).
\bibitem{Bassi05} A. Bassi, {\it Collapse models: analysis of the free particle dynamics}, J. Phys. A {\bf 38}, 3173 (2005).
\bibitem{Adler07} S. L. Adler and A. Bassi, {\it Collapse Models with Non-White Noises}, J. Phys. A {\bf 40}, 15083 (2007).
\bibitem{Adler08} S. L. Adler and A. Bassi, {\it Collapse models with non-white noises: II. Particle-density coupled noises}, J. Phys. A {\bf 41}, 395308 (2008).
\bibitem{Bassi13} A. Bassi, K. Lochan, S. Satin, T. P. Singh, and H. Ulbricht, {\it Models of wave-function collapse, underlying theories, and experimental tests}, Rev. Mod. Phys. {\bf 85}, 471 (2013).

\bibitem{Vinante16} A. Vinante, M. Bahrami, A. Bassi, O. Usenko, G. Wijts, and T. H. Oosterkamp,
{\it Upper Bounds on Spontaneous Wave-Function Collapse Models Using Millikelvin-Cooled Nanocantilevers}, Phys. Rev. Lett. {\bf 116}, 090402 (2016).
\bibitem{Vinante17} A. Vinante, R. Mezzena, P. Falferi, M. Carlesso, and A. Bassi, {\it Improved Noninterferometric Test of Collapse Models Using Ultracold Cantilevers}, Phys. Rev. Lett. {\bf 119}, 110401 (2017).
\bibitem{Bahrami18} M. Bahrami, {\it Testing collapse models by a thermometer}, Phys. Rev. A {\bf 97}, 052118 (2018).
\bibitem{Tilloy19} A. Tilloy and T. M. Stace, {\it Neutron Star Heating Constraints on Wave-Function Collapse Models}, Phys. Rev. Lett. {\bf 123}, 080402 (2019).
\bibitem{Pontin19} A. Pontin, N. P. Bullier, M. Toro\v{s}, and P. F. Barker, {\it Ultranarrow-linewidth levitated nano-oscillator for testing dissipative wave-function collapse}, Phys. Rev. Res. {\bf 2}, 023349 (2020).
\bibitem{Vinante20} A. Vinante, M. Carlesso, A. Bassi, A. Chiasera, S. Varas, P. Falferi, B. Margesin, 
R. Mezzena, and H. Ulbricht, {\it Narrowing the Parameter Space of Collapse Models with Ultracold Layered Force Sensors}, Phys. Rev. Lett. {\bf 125}, 100404 (2020).
\bibitem{Zheng20} D. Zheng, Y. Leng, X. Kong, R. Li, Z. Wang, X. Luo, J. Zhao, C.-K. Duan, P. Huang, 
J. Du, M. Carlesso, and A. Bassi, {\it Room temperature test of the continuous spontaneous localization model using a levitated micro-oscillator}, Phys. Rev. Res. {\bf 2}, 013057 (2020).
\bibitem{Komori20} K. Komori, Y. Enomoto, C. P. Ooi, Y. Miyazaki, N. Matsumoto, V. Sudhir, Y. Michimura, 
and M. Ando, {\it Attonewton-meter torque sensing with a macroscopic optomechanical torsion pendulum}, Phys. Rev. A {\bf 101}, 011802(R) (2020).
\bibitem{Donadi21} S. Donadi, K. Piscicchia, C. Curceanu, L. Di\'{o}si, M. Laubenstein, and A. Bassi, {\it Underground test of gravity-related wave function collapse}, Nat. Phys. {\bf 17}, 74 (2021).
\bibitem{Gasbarri21} G. Gasbarri, A. Belenchia, M. Carlesso, S. Donadi, A. Bassi, R. Kaltenbaek,
M. Paternostro, and H. Ulbricht, {\it Testing the foundation of quantum physics in space via Interferometric and non-interferometric experiments with mesoscopic nanoparticles}, Commun. Phys. {\bf 4}, 155 (2021).
\bibitem{Carlesso22} M. Carlesso, S. Donadi, L. Ferialdi, M. Paternostro, H. Ulbricht, and A. Bassi, {\it Present status and future challenges of non-interferometric tests of collapse models}, Nat. Phys. {\bf 18}, 243 (2022).

\bibitem{Myrvold17} W. C. Myrvold, {\it Relativistic Markovian dynamical collapse theories must employ nonstandard degrees of freedom}, Phys. Rev. A {\bf 96}, 062116 (2017).
\bibitem{Tumulka20} R. Tumulka, {\it A Relativistic GRW Flash Process with
Interaction} (Springer, New York, 2020).
\bibitem{Jones20} C. Jones, T. Guaita, and A. Bassi, {\it Impossibility of extending the Ghirardi-Rimini-Weber model to relativistic particles}, Phys. Rev. A {\bf 103}, 042216 (2021).
\bibitem{Jones21} C. Jones, G. Gasbarri, and A. Bassi, {\it Mass-coupled relativistic spontaneous collapse models}, J. Phys. A: Math. Theor. {\bf 54}, 295306 (2021).

\bibitem{Wang22} P. Wang, {\it Relativistic quantum field theory of stochastic dynamics in the Hilbert space}, Phys. Rev. D {\bf 105}, 115037 (2022).
\bibitem{Wang24} P. Wang, {\it Relativistic model of spontaneous wave-function localization induced by nonHermitian colored noise}, arXiv: 2501.07050.

\end{thebibliography}
\end{document}